\documentclass[aps,prl,reprint,superscriptaddress,showpacs]{revtex4-1}
\usepackage[T1]{fontenc}
\usepackage[utf8]{inputenc}
\usepackage{color}
\usepackage{graphicx}
\usepackage{amsmath}
\usepackage{amssymb}
\usepackage{amstext}
\usepackage{amsthm}
\usepackage{latexsym}
\usepackage{verbatim}
\usepackage{natbib}
\usepackage{array}

\usepackage{ulem}   

\begin{document}

\title {Formation of an incoherent metallic state in Rh-doped Sr$_2$IrO$_4$}
\vspace{1cm}

\author{A. Louat}

\author{F. Bert}

\author{L. Serrier-Garcia}
\affiliation {Laboratoire de Physique des Solides, CNRS, Univ. Paris-Sud, Universit\'{e} Paris-Saclay, 91405 Orsay Cedex, France}

\author{F. Bertran}

\author{P. Le F\`{e}vre}

\author{J. Rault}
\affiliation {Synchrotron SOLEIL, L'Orme des Merisiers, Saint-Aubin-BP 48, 91192 Gif sur Yvette, France}

\author {V. Brouet}
\affiliation {Laboratoire de Physique des Solides, CNRS, Univ. Paris-Sud, Universit\'{e} Paris-Saclay, 91405 Orsay Cedex, France}

\begin{abstract}

Sr$_2$IrO$_4$ is the archetype of the spin-orbit Mott insulator, but the nature of the metallic states that may emerge from this type of insulator is still not very well known. We study with angle-resolved photoemission the insulator-to-metal transition observed in Sr$_2$Ir$_{1-x}$Rh$_x$O$_4$ when Ir is substituted by Rh (0.02 < $x$ < 0.35). The originality of the Rh doping is that Ir and Rh, which are formally isovalent, adopt different charge states, a rather unusual and inhomogeneous situation. We show that the evolution to the metallic state can be essentially understood as a shift of the Fermi level into the lower Hubbard band of Sr$_2$IrO$_4$. The Mott gap appears quite insensitive to the introduction of up to $\sim$20\% holes in this band. The metallic phase, which forms for $x$ > 0.07, is not a Fermi liquid. It is characterized by the absence of quasiparticles, unrenormalized band dispersion compared to calculations and an $\sim$30-meV pseudo-gap on the entire Fermi surface. 
\end{abstract}

\date{\today}

\maketitle

\lq\lq{}Bad metallic states\rq\rq{} are often observed in the vicinity of a metal-insulator transition and may lead to non-Fermi-liquid behaviors. The description of this intriguing state of matter is difficult, as disorder and correlations may both play an essential role \cite{dobrosavljevic2012introduction}. Introducing carriers in a Mott insulator destabilize the Mott state, but typically also introduce disorder. Disentangling the two effects is usually very difficult. We study here the case of Rh doped Sr$_2$IrO$_4$, which appears as a particularly interesting example of this competition. Starting from the celebrated spin-orbit Mott insulating state in Sr$_2$IrO$_4$ \cite{BJKimPRL08}, Rh substitutions are able to induce an insulator to metal transition around $x\sim$0.07 \cite{QiCaoPRB12}, by adding hole carriers \cite{ClancyPRB14,CaoNatCom16}. They are also creating strong disorder, as Rh directly substitutes for Ir on the active IrO$_2$ plane and adopts a different charge state than Ir. The optimal metallic state is reached for $x\sim$0.15 \cite{QiCaoPRB12,BrouetPRB15}. The resistivity increases again upon further doping, which was assigned to Anderson localization \cite{QiCaoPRB12}, making it clear that the role of disorder cannot be ignored. 

Iridium has a valence of 4+ in Sr$_2$IrO$_4$, meaning five electrons in the 5d t$_{2g}$ orbitals. Surprisingly, x-ray absorption measurements have shown that Rh is close to the 3+ valence state at a small $x$ value, implying that it has trapped one electron from the surrounding Ir ions \cite{ClancyPRB14,ChikaraPRB17}. Consequently, Rh induces a hole doping on the Ir sites, which has indeed been confirmed by an angle-resolved photoemission spectroscopy (ARPES) study \cite{CaoNatCom16}. The reason why Rh takes a different valence than Ir has not really been clarified so far. A major difference between Ir and Rh is the much smaller spin-orbit coupling (SOC) of Rh. It was suggested in Ref. \cite{CaoNatCom16} that this alone could favor a larger occupation of Rh ions. However, this is based on a simple atomic picture. The situation in the solid is far more complex and interesting. It just starts to be explored by first-principle calculations \cite{LiuPRB16}. The concept of the spin-orbit Mott insulator is that the Mott insulating state will appear when the degeneracy of the t$_{2g}$ orbitals is completely lifted by a strong SOC, promoting a narrow half-filled band at the Fermi level \cite{BJKimPRL08,MartinsPRL11}. This occurs for Sr$_2$IrO$_4$ \cite{CrawfordPRB94} and not for the isostructural Sr$_2$RhO$_4$, which remains metallic \cite{PerryNJP06}. Could this smaller effective correlation on Rh play a role in the distribution of electrons among Ir and Rh ? How would the Mott gap on Ir be affected by this type of hole doping ?  


We present an ARPES study covering a large range of Rh dopings to follow in detail the behavior of the SOC splitted t2g bands across the successive insulator-to-metal and metal to insulator transitions. We find that a rigid band filling picture describes the evolution to a very large extent, with a continuous shift at early dopings and no closure of the Mott gap by more than 20\% at 15\% hole doping. In the metallic state, we note the absence of well-defined quasiparticle peaks (QPs) and the presence of an $\sim$30-meV pseudogap on the entire Fermi surface (FS). This pseudogap is however different both from the one found in the cuprates and from the one reported in Ref. \cite{CaoNatCom16}. We assign it to disorder effects associated with Rh substitutions in the correlated insulator. We rationalize the presence of metallic or insulating regions by the Mott-Ioffe-Regel criterion \cite{MottReview87}, comparing the mean free path $l$ and the average distance between carrier $d$, both determined from our ARPES measurements. We observe a saturation of the number of holes carriers at high Rh doping not observed previously \cite{CaoNatCom16} and in good agreement with Ref. \cite{ChikaraPRB17}.

\begin{figure*}[tb]
\centering
\includegraphics[width=0.9\textwidth]{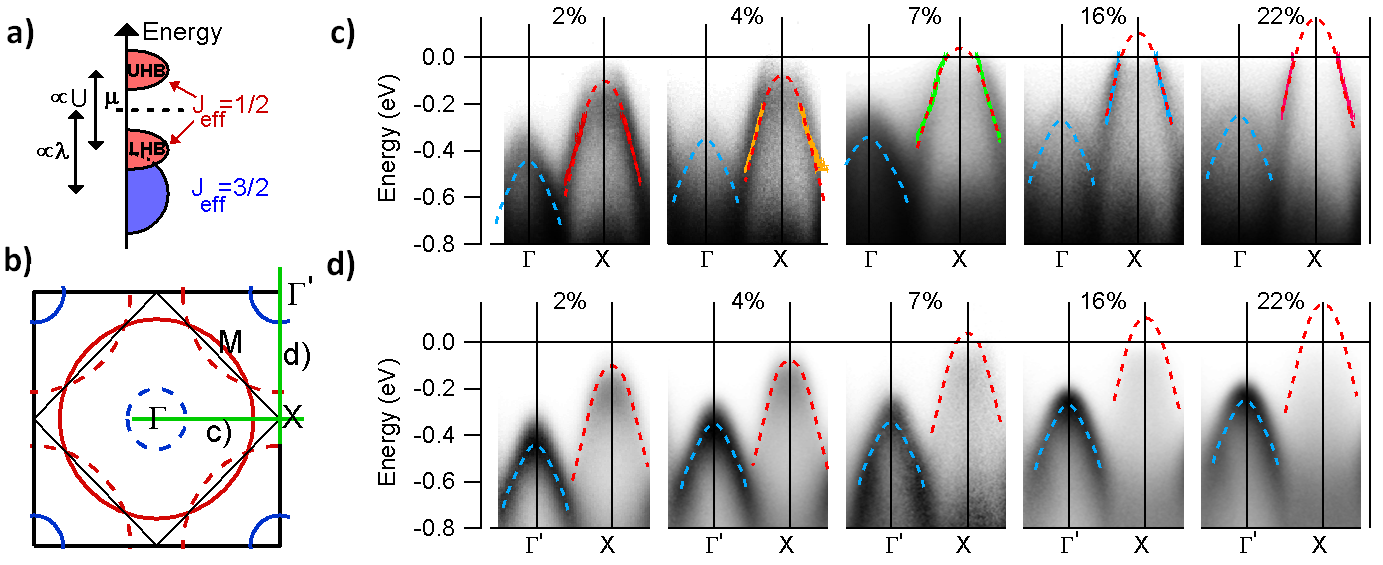}
\caption{ (a) Sketch of the bands expected in Sr$_2$IrO$_4$ as a function of energy. The J$_{1/2}$ band is divided into lower Hubbard band (LHB) and the upper Hubbard band (UHB). The chemical potential $\mu$ is represented by a dashed black line. $U$ is the Coulomb repulsion, and $\lambda$ is the SOC constant. (b) Sketch of the Fermi surface calculated for Sr$_2$IrO$_4$. The thin black square corresponds to the 2Ir Brillouin zone (BZ) and the dotted lines are obtained by folding with respect to these 2Ir BZ boundaries \cite{sup}. The green lines indicate the cuts plotted in (c) and (d). (c) Energy-momentum plots at 50K along $\Gamma X$ for the indicated Rh dopings. The J$_{1/2}$ and J$_{3/2}$ bands are underlined by red and blue guides to the eye, respectively. The symbols indicate the dispersion of J$_{1/2}$ extracted by fits of the Momentum Distribution Curves (MDC). (d) Energy-momentum plots along $\Gamma ' X$, where J$_{3/2}$ is clearer.}
\label{Fig1}
\end{figure*}

The samples were prepared using a self-flux method, as reported in Ref. \cite{KimScience09}. Their exact doping was estimated by energy dispersion x-ray analysis, and the basic evolution of the resistivity and magnetization is shown in the Supplementary Material \cite{sup}. ARPES experiments were carried out at the CASSIOPEE beamline of the SOLEIL synchrotron, with a SCIENTA R-4000 analyzer and an overall resolution better than 15meV. All data shown here were acquired at a photon energy of 100eV, with linear polarization and at a temperature of 50K.

\vspace{0.5cm}  


The electronic structure of Sr$_2$IrO$_4$ is well known from previous ARPES studies \cite{BJKimPRL08,WangDessauPRB13,DelaTorrePRL15,BrouetPRB15}. The SOC splits the t2g levels into a band with J$_{eff}$=3/2 (J$_{3/2}$) character filled with four electrons and a half-filled J$_{eff}$=1/2 (J$_{1/2}$) band, further split by correlation effects [see sketch of Fig. 1(a)]. The full non-interacting band structure is given in the Supplementary Material \cite{sup} and the sketch of the corresponding Fermi surface reported in Fig. 1(b). The solid lines in Fig. 1(b) describe the Fermi surface that would be expected for a one Ir Brillouin zone (the thick black square). In reality, there are two Ir in the unit cell, inequivalent because of in-plane rotation of oxygen octahedra \cite{CrawfordPRB94}, and all lines are folded into the 2 Ir BZ (the thin black square). We show them as the dotted lines, as their intensity is typically weaker in ARPES \cite{WangDessauPRB13}. This folding is usually absent in cuprates and will be important for our discussion of Fermi pockets or Fermi arcs appearing upon doping.

\begin{figure}[b]
\centering
\includegraphics[width=0.48\textwidth]{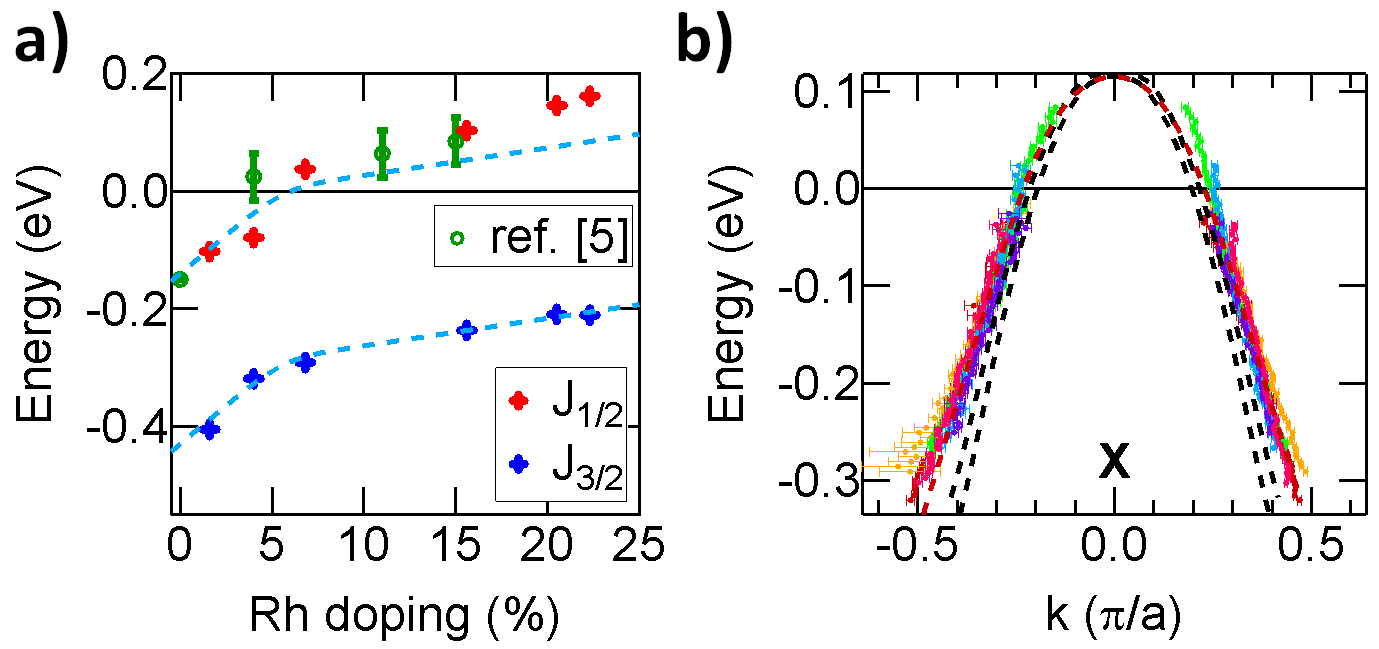}
\caption{ (a) Position of the top of the J$_{1/2}$ and J$_{3/2}$ models shown in Fig.1. The bottom blue dashed line is a guide to the eye for the J$_{3/2}$ shift. It is reproduced with a vertical shift to match the J$_{1/2}$ data at small doping. Data from Ref. \cite{CaoNatCom16} are added as the green open circles. b) Superposition of the MDC dispersion for J$_{1/2}$, displayed in Fig.1 (the colors are kept the same). They are shifted to the dispersion calculated for Sr$_2$IrO$_4$ (the black dashed lines, there are two bands due to inequivalent Ir in the unit cell). }
\label{shift}
\end{figure}

In Figs.1(c) and 1(d), we show the evolution of the two bands with Rh doping, along the two perpendicular directions $\Gamma$'X and $\Gamma$X. We highlight the dispersion of J$_{3/2}$ with a blue parabola and that of J$_{1/2}$ with a red one. These guides to the eyes are defined for $x$=0.02 and only shifted up as a function of Rh doping. This illustrates the evolution is essentially a shift towards the Fermi level, without changes in the shapes of the dispersions. The top positions of these guides are reported in Fig. \ref{shift}(a). There is no band crossing the Fermi level up to $x\sim$ 4\%, as expected for the insulating state. For doping of about 7\%, the $J_{1/2}$ band reaches the Fermi level, which corresponds well to the insulator-metal transition observed in transport \cite{QiCaoPRB12}. At higher doping, hole pockets form around X. 

\begin{figure*}[tb]
\centering
\includegraphics[width=1\textwidth]{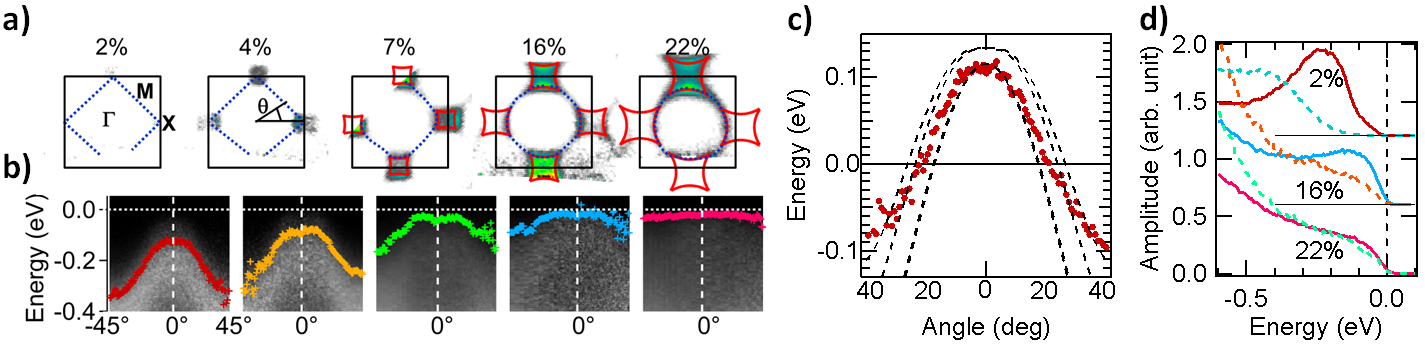}
\caption{ a) Maps of the spectral weight integrated in a 10~meV window around E$_F$ for different Rh dopings at 50K. The dotted blue line indicates for each $\theta$ value the point where the spectrum is closest to E$_F$. The red lines indicate the hole pockets (see the text). (b) Map of the energy distribution curve (EDC) closest to E$_F$ as a function of angle $\theta$. The crosses indicate the leading edge of the EDC. (c) The leading edge dispersion at $x$ = 2\% shifted up by 0.235~eV and compared to the dispersion calculated along XM for Sr$_2$IrO$_4$. (d) EDC spectra showing the leading edge along $\Gamma$X (red, blue, and purple) and $\Gamma$M in the complementary color, for the indicated dopings. }
\label{FS}
\end{figure*}

We also report in Fig.\ref{shift}(a), with green symbols, results from the previous ARPES study \cite{CaoNatCom16}. They are in overall good agreement, although it was argued in Ref. \cite{CaoNatCom16} that the Fermi level jumps between 0 and 4 \%. In our case, the evolution is clearly continuous, which could be due to a residual density of states within the gap or an intrinsic change in $\mu$ \cite{CamjayiPRB06}. As there are no points between $x$=0 and 0.04 in the data of Ref. \cite{CaoNatCom16}, it is difficult to decide if the behavior is qualitatively different or not. 

To better understand the meaning of this shift, it is instructive to compare the evolution of J$_{3/2}$ and J$_{1/2}$. Fig.\ref{shift}(a) shows that the distance between them remains remarkably stable with doping. To evidence this, we report the variation of J$_{3/2}$ (the dashed blue line) at the position of the J$_{1/2}$. They are similar except for x > 10\%, where the shift is stronger for J$_{1/2}$ by $\sim$50 meV. As recalled in Fig. 1(a), this distance is mainly controlled by the relative strength of spin-orbit coupling $\lambda$ and the Coulomb repulsion $U$. 
If the first effect of Rh substitution was a reduction of SOC, as originally expected \cite{LeeTokuraPRB12}, one would expect this distance to decrease. 
On the other hand, if the gap within the J$_{1/2}$ band was decreasing, one would expect an increase in the distance between J$_{3/2}$ and J$_{1/2}$. This is the most natural explanation for the small increase observed in the metallic region. The gap in Sr$_2$IrO$_4$ is evaluated to about 0.6eV in optical \cite{MoonPRB09}, scanning tunneling microscopy (STM) \cite{BattistiNatPhys17}, or ARPES measurements \cite{BrouetPRB15}. This suggests that the gap is only marginally reduced by about 0.1 eV and that metallicity is essentially induced by creating holes in the lower Hubbard band. 

To check more precisely the validity of the rigid band scenario, we superpose in Fig.\ref{shift}(b), all the Momentum Distribution Curves (MDCs) dispersions between $x$ = 0.02 and 0.22. The excellent overlap proves, within experimental accuracy, that there is no sizable deviation from a rigid band shift. Comparison with the dispersion calculated for the J$_{1/2}$ band [black dashed lines in Fig. 2(b)] shows there is no need for renormalization. In fact, there is no change in the band shape when the system becomes metallic, suggesting that the nature of this band does not change and remains essentially incoherent.

\vspace{0.5cm}

In Fig. \ref{FS}, we describe the hole pockets more precisely, by mapping the Fermi surface in the full BZ. Squarish hole pockets emerge around X (the red contours) with increasing Rh content. 
To define properly this FS, we track the leading edge of the spectra as a function of the angle $\theta$ [0$^{\circ}$ corresponds to X and 45$^{\circ}$ to M, see Fig. \ref{FS}(a)]. For each $\theta$ direction, we extract the EDC closest from E$_F$. The dotted blue line in Fig. \ref{FS}(a) tracks the location of these EDCs at each $\theta$. This contour evolves from nearly a square at low dopings to a circle at larger dopings. We build the color image of Fig. \ref{FS}(b) from these EDCs. The crosses in Fig.\ref{FS}(b) locate the half maximum of the leading edge of the EDC. The band is closest to E$_F$ at X and shifts up with doping. Nevertheless, as shown in Fig. \ref{FS}d, the leading edge never reaches zero, as would be expected for a good metal, but saturates around $\sim$30 meV, which we will call a pseudogap. The red contours in Fig. \ref{FS}(a) delimit the positions at which this value is reached for all doping levels. For $x$ = 0.22, this is almost a circle with no $k$-dependence of the pseudogap. However, the corresponding EDCs still lack a well defined QP peak, which is another evidence for the incoherent nature of the excitations.

We use here the term "pseudogap" to describe the lack of density of states at E$_F$, irrespective of its origin. Inevitably, this questions a possible relation to cuprates. Hole-doped iridates should be compared to electron-doped cuprates, because of an opposite band curvature \cite{WangSenthilPRL11}. At early dopings, the FS also develops around X in electron-doped cuprates \cite{ArmitagePRL02}, but through the form of Fermi arcs, as there is no folding due to a doubling of the unit cell [i.e., no FS corresponding to the dotted lines in Fig. 1-b)]. The lack of intensity at M can be viewed as a pseudogap. For further doping ($x\sim$0.15, or even as early as $x\sim$0.05 \cite{SongPRL17}), an electron pocket develops around M, formed by the UHB when the gap closes. This situation is not observed in Rh doped iridates, in line with our previous suggestion that the gap is still far from closing, even for 20\% Rh. In our case, the difference in position between X and M can be simply understood from the dispersion expected along this direction. We show in Fig. \ref{FS}(c) that the calculated dispersion along XM fits well with the leading edge dispersion obtained at low dopings, where it almost follows XM. As the doping increases, the band shifts up rigidly, without changing its spectral lineshape notably as a function of $k$. It first crosses at X point, leaving a residual gap at M, rather then a pseudogap. 

The pseudogap we discuss here occurs on the FS sections, which would be metallic in electron-doped cuprates. In Ref. \cite{CaoNatCom16}, similar hole pockets were reported around X, but with a pseudogap only on the parts of the pockets corresponding to the solid lines of Fig. 1(b), not on the folded ones. Although this would form arcs, this is different from the case of the cuprates, and the breaking of symmetry that could cause such a difference is very hard to conceive, as discussed in this paper. We do not observe such a difference between folded and non-folded sheets and believe it may be caused by matrix element effects \cite{FuturRh}. As our pseudogap is not $k$-dependent and associated with the nearly metallic parts of the electronic structure, we take it as a fingerprint of the disordered metallic state reached through Rh substitutions and continue with the characterization of this state.    
 
\begin{figure}[tb]
\centering
\includegraphics[width=0.48\textwidth]{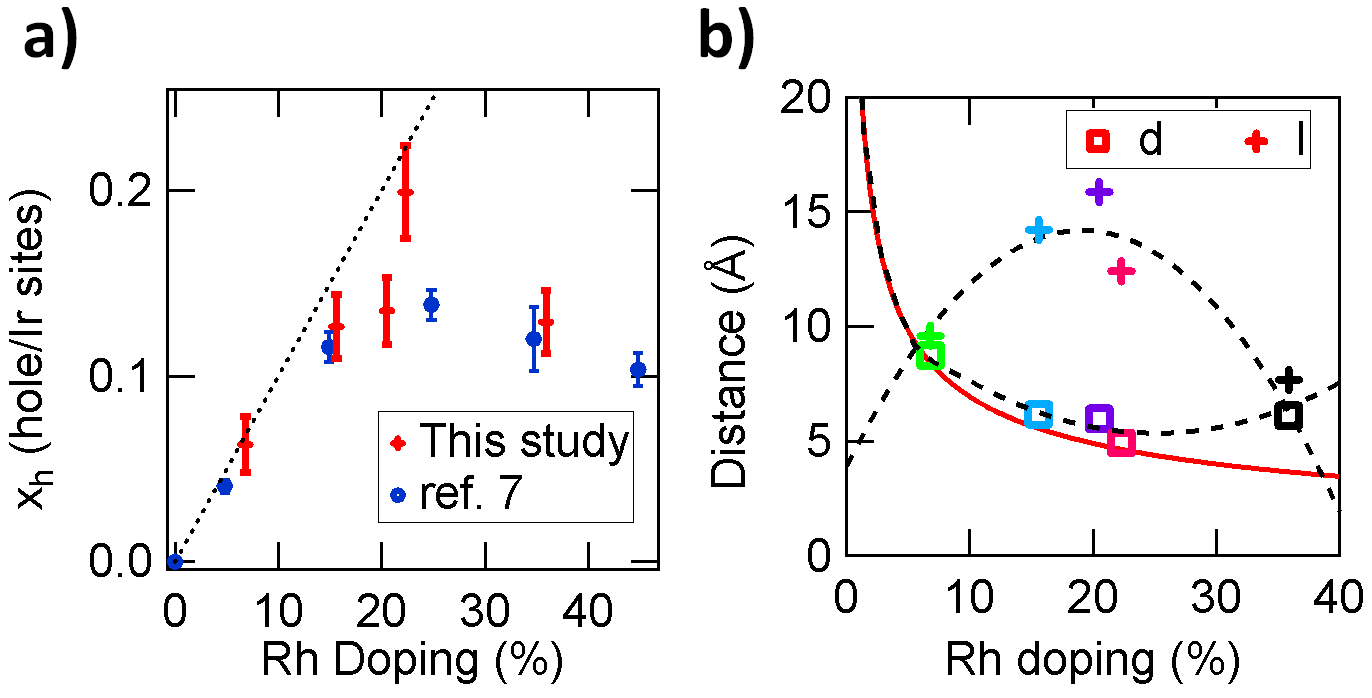}
\caption{ (a) The red symbols : the number of holes $x_h$ as a function of Rh doping $x_{Rh}$, computed from the area of the hole pockets shown in Fig.3. The blue symbols : Rh valence extracted from x-ray absorption spectroscopy (XAS) measurements \cite{ChikaraPRB17}. (b) The cross : mean free path $l$ computed from the MDC width. Open square : average distance between carriers $d$, deduced from $x_h$ (see the text). The red line is the value expected for $x_h$=$x_{Rh}$. The black dashed lines are guides to the eye. }
\label{Final}
\end{figure}
 
Taking the red contours of Fig. \ref{FS}(a) as a measure of the FS, we can compute the number of holes $x_h$ as a function of Rh doping from the Luttinger theorem \cite{LuttingerPR60}. The area of the hole pocket, $A_{pocket}$, is roughly given by that of a square of width 2$k_F$. More precisely, it is defined by the intersection of circles of radius (1-k$_F$) centered at $\Gamma$ and $\Gamma'$ \cite{sup}. As there is one such pocket per 2 Ir BZ, we obtain, using $x_h$=2$A_{pocket}$/$A_{BZ}$, the values plotted as the red points in Fig. \ref{Final}(a). Interestingly, this follows very well the Rh doping at early values, as if each Rh trapped one electron and gave one hole to Ir, consistent with the 3+ valence state found in XAS \cite{ClancyPRB14}. We further find at larger dopings a saturation of the number of holes, indicating the additional Rh electron is partially released. This finding is in excellent agreement with a recent XAS study, which obtained values reported as the blue circles in Fig. \ref{Final}(a) \cite{ChikaraPRB17}.

We can go further into the nature of the metal by looking at the MDC width at E$_F$, displayed in Fig. \ref{Final}(b). Neglecting all other sources of broadening, it gives an estimate of the mean free path $l\sim1/\Delta\nu$. Around x$_{Rh}$=20\%, a maximum value of $l$ = 15\AA~is found, which is short, but longer than the distance between Ir, $a\sim$4\AA, suggesting metallic conduction can exist. More precisely, the Mott-Regel-Ioffe criterion compares $l$ with the average distance between carriers $d$, expecting a metal for $l$ > $d$ (the exact numerical value for the metal-insulator transition can be discussed \cite{GrahamJPhysCondMat98}). $d$ can be deduced from our measurement of $k_F$, as $k_F$ directly defines the number of carriers $x_h$ [see Fig. 4(a)]. Assuming these carriers are randomly distributed on the square lattice, we expect $d=a/\sqrt{\pi x_{h}}$ \cite{sup}. In the insulating region, we cannot define $k_F$, but we can estimate $d$ using $x_h$=$x_{Rh}$ (the red line), which holds at low dopings [Fig. 4(a)]. In Fig. \ref{Final}(b), we compare $l$ to $d$ and find that the condition $l$ > $d$ will not be fulfilled at low and high Rh dopings. At high doping, the disorder induced by Rh on the IrO$_2$ plane decreases the mean free path $l$ to the point that it becomes smaller than $d$. This is in line with the idea of Anderson localization at high Rh doping deduced from transport measurements \cite{QiCaoPRB12}. At low dopings, there are few holes, so that the distance between them becomes longer than $l$. 

\vspace{0.5cm}

An interesting question raised by this study is why the metallic state exhibits a pseudogap. Although there would be a finite density of states at $E_F$ for a purely disorder-driven Anderson insulator, including long range electronic correlations would open a soft Coulomb gap on the insulating side of the transition \cite{EfrosShklovskii75} and corrections to the density of states mimicking a pseudogap on the metallic side \cite{AltshulerAronov83}. More recently, the effect of short-range Coulomb interactions, which is most relevant for this correlated systems, has been investigated theoretically, and soft Hubbard gaps were predicted \cite{ShinaokaPRL09,LeeValentiPRB16}. Experimentally, disordered correlated systems are rare \cite{LahoudPRL14} and this system offers a good opportunity to test this behavior.     


Beyond a structural disorder, Rh creates charged defects, which seems a rather strong perturbation. It is interesting to observe that iridates tolerate or even promote such defects. This may also have implications for other types of defects (for example, local oxygen defects, such as distortion and vacancy, which are common in oxides) or doping, even for La/Sr substitutions. What type of metallic state emerges from this situation is a novel question. Our ARPES study suggests the metallic behavior upon Rh doping essentially develops in the lower Hubbard band, whereas the Mott gap is not strongly affected. This would defer from the case of La substitutions, where a collapse of the Mott gap was reported for 8\% electron doping \cite{DelaTorrePRL15,BattistiNatPhys17}. Of course, a direct measurement of the full gap, for example, through scanning tunneling spectroscopy measurements, would be desirable to complement this picture. The absence of sensitivity of the Mott gap may be linked to the fact that there is no real emergence of a coherent behavior. The dispersions are unrenormalized compared to band calculation and a pseudogap develops over the entire Fermi surface.

\vspace{0.5cm}

We thank M. Civelli, S. Biermann and C. Martins for useful discussions. We thank ANR \lq\lq{}SOCRATE\rq\rq{} (Grant No. ANR-15-CE30-0009-01), the Universit\'e Paris-Sud (\lq\lq{}PMP MRM Grant\rq\rq{}) and the Investissement d’Avenir LabEx PALM (Grant No. ANR-10-LABX-0039-PALM) for financial support.

\bibliography{Rhpaper_biblio}

\newpage
\begin{widetext}
\Large
\textbf{Supplementary Information}
\vspace{0.4cm}

\noindent\normalsize
\noindent

\textbf{Sample resistivity and magnetization}

The magnetization and resistivity measurements were carried out using a commercial Superconducting Quantum Interference Device Magnetometer (Quantum Design MPMS SQUID) and a commercial Physical Properties Measurement System (Quantum Design PPMS). For large Rh doping (about 15\%) the resistivity slope for temperature above 100~K is positive. Above this optimal doping, the slope becomes negative again.

\begin{figure*}[h]
\centering
\includegraphics[width=0.8\textwidth]{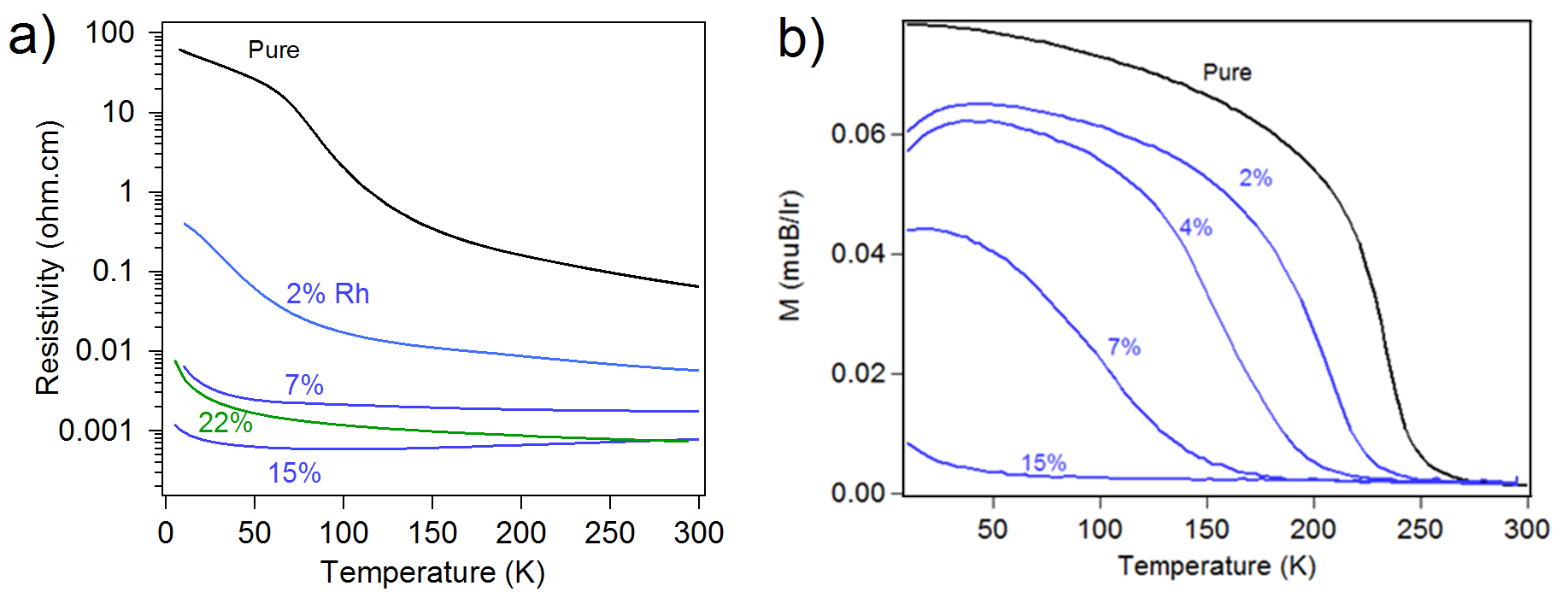}
\caption{Evolution of a) the resistivity and (b) the magnetization at 1~T as a function of the temperature for Sr$_2$IrO$_4$ doped with Rh.}
\label{Sup3}
\end{figure*}

\noindent
\textbf{LDA calculation}

We reproduce here the band structure calculation we use in the manuscript. It is similar to previous calculations \cite{BJKimPRL08,MartinsPRL11}. Although the band structure is essentially two dimensional, there are small differences between $k_z$=0 and $k_z$=1 and we have used the bands at $k_z$=1 as a reference. The structure hardly changes with Rh doping \cite{YeCaoPRB13}, so that changes of the band structure itself can be neglected as a first approximation.

In the Fermi Surface corresponding to this calculation, the J=3/2 band is expected to cross the Fermi level to form a small hole pocket (in blue) around $\Gamma$. In experiment, it is pushed at -0.4eV below the Fermi level, which is understood from correlation enhanced spin-orbit coupling. Compared to the atomic values $\lambda$=0.36eV for Ir and 0.13eV for Rh, correlation effects increase these values by a factor 2, both for Ir \cite{MartinsPRL11,zhou2017correlation} and Rh \cite{LiuPRL08}.

\begin{figure*}[h]
\centering
\includegraphics[width=0.8\textwidth]{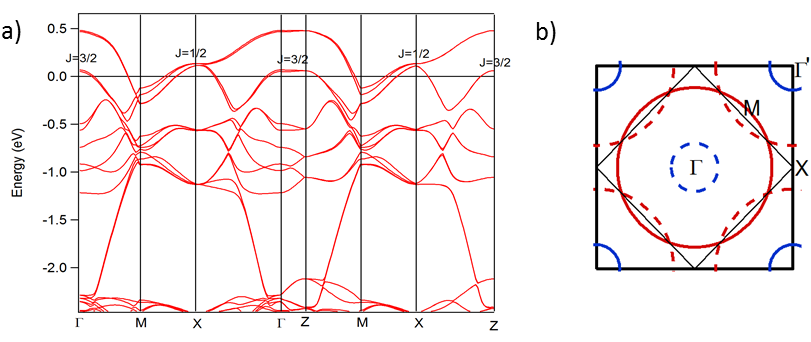}
\caption{a) Band calculations obtained using the Wien2k package \cite{Wien2k} with the experimental structure of Sr$_2$IrO$_4$. There are 4 inequivalent Ir in the unit cell, 2 in-plane because of the different oxygen octahedra rotations in-plane \cite{DhitalPRB13} (they are also inequivalent with respect to AF order) and 2 inequivalent planes. The main character of the bands crossing the Fermi level are indicated. b) Sketch of the Fermi Surface corresponding to this band calculation. }
\label{Fig1}
\end{figure*}

\vspace{0.4cm}
\newpage
\noindent
\textbf{Conversion from $k_F$ to hole pocket area}

For a Fermi wave vector $k_F$ between $0$ and $1-1/\sqrt{2}$, the hole pocket area in the BZ ($A$) is given by
\begin{equation}
A = 8 \left[\left(1-(1-k_F) \cos(\theta) \right)^2 - (1-k_F) \left(\theta - \frac{1}{2} \sin(2\theta) \right) \right]
\end{equation}

$\theta$ is the angle between $\Gamma X$ direction and the intersection point between circles (J$_{1/2}$). $\theta(k_F) = \frac{\pi}{4} - \arccos\left(\frac{1}{\sqrt{2} (1-k_F)}\right)$

\begin{figure*}[h]
\centering
\includegraphics[width=0.5\textwidth]{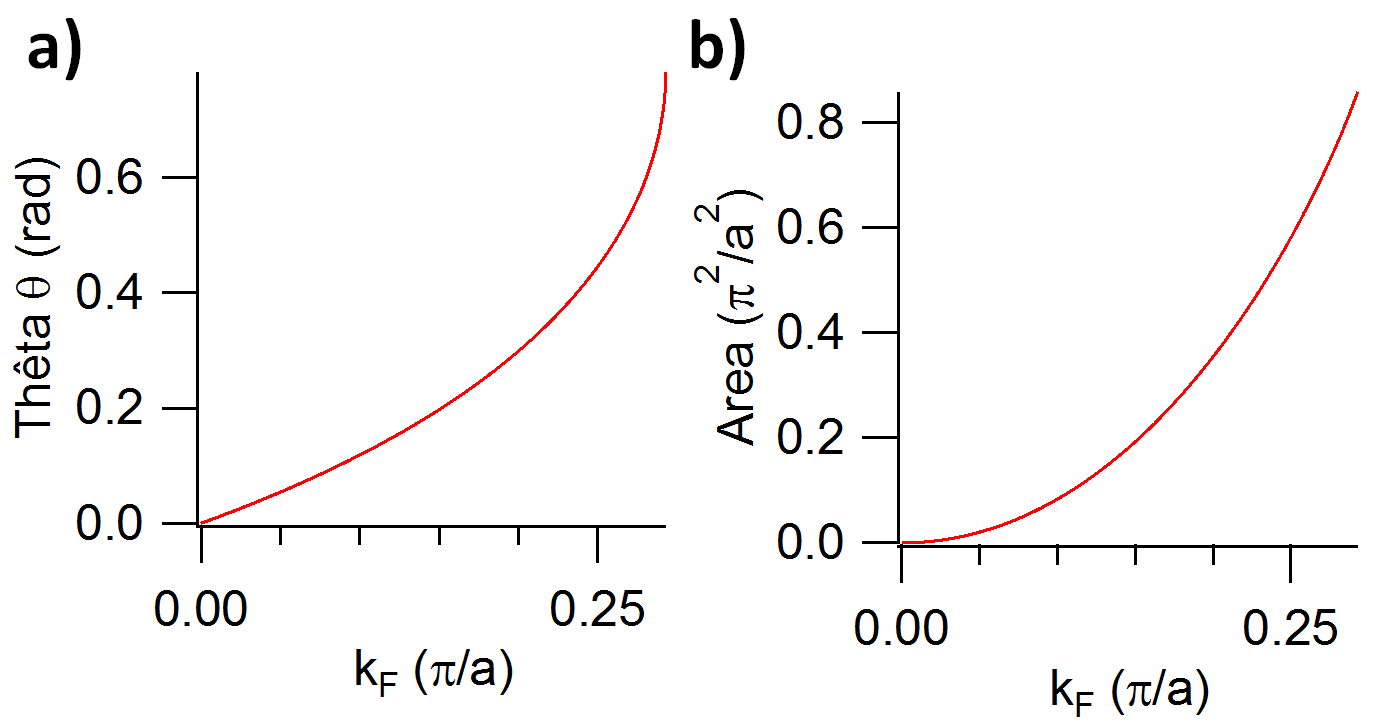}
\caption{a) Variation of the angle $\theta$ between the $\Gamma X$ direction and the cross between circles formed by J$_{1/2}$ band and b) variation of hole pocket area in the first BZ as a function of the Fermi wave vector $k_F$ between $1/\sqrt{2}$ and 1. }
\label{Fig1}

\end{figure*}

From the number of holes, we deduce the average distance between them $d$. We assume a random distribution, so that there are $x_h$ holes per unit square of surface a$^2$. The area covered by each hole is $\pi$d$^2$, giving $d=a/\sqrt{\pi x_{h}}$.
\end{widetext}
\end{document}